\documentclass[pra,letterpaper,aps,twocolumn]{revtex4}
\usepackage{graphicx}
\usepackage{amsmath}

\newcommand{\be}{\begin{equation}}
\newcommand{\ee}{\end{equation}}
\newcommand{\bk}{{{\bf{k}}}}

\newcommand{\cE}{{{\cal E}}}
\newcommand{\bQ}{{{\bf{Q}}}}
\newcommand{\br}{{{\bf{r}}}}

\newcommand{\bg}{{{\bf{g}}}}

\newcommand{\bR}{{{\bf{R}}}}

\newcommand{\bea}{\begin{eqnarray}}
\newcommand{\eea}{\end{eqnarray}}
\newcommand{\ra}{\rangle}
\newcommand{\la}{\langle}
\newcommand{\upa}{\uparrow}
\newcommand{\dna}{\downarrow}

\newcommand{\dg}{{\dagger}}
\newcommand{\pdg}{{\phantom\dagger}}
\begin{document}
\title{Multiband superfluidity and superfluid to band-insulator transition
of strongly interacting fermionic atoms in an optical lattice}
\author{A.A. Burkov}
\affiliation{Department of Physics and Astronomy, University of Waterloo, 
Waterloo, Ontario, Canada N2L 3G1}
\author{Arun Paramekanti}
\affiliation{Department of Physics, University of Toronto, Toronto, Ontario, 
Canada M5S 1A7}  
\date{\today}
\begin{abstract}
We study the multiband superfluid and the superfluid (SF) to band
insulator (BI) transition
of strongly interacting fermionic atoms in an 
optical lattice at a filling of two fermions per well. We present physical
arguments to show that a
consistent mean field description of this problem is obtained by
retaining only intraband pairing between the fermions. Using this approach we obtain
a reasonable account of the experimentally observed
critical lattice depth for the SF-BI transition and the modulated 
components of the condensate density, and make predictions for 
the lattice depth dependence of the quasiparticle gap which can be tested in
future experiments.
We also highlight
some interesting features unique to cold atom superfluids within this
intraband pairing approximation; for instance,
the pair field is forced to be uniform in space and the Hartree field
vanishes identically. These arise as a result of the fact that while the pairing
interaction is cut off at the scale of the Debye frequency in conventional
superconductors, or at the lattice scale in tight binding model Hamiltonians,
such a cutoff is absent for cold Fermi gases.
\end{abstract}
\maketitle

\section{Introduction} 
Studies of cold atomic gases have
led to significant experimental 
\cite{bcsbec-expt}
and theoretical progress 
\cite{bcsbec-theory}
in our understanding
of the crossover regime between a weakly paired BCS superfluid and 
a strongly paired BEC superfluid. Recent experiments have also studied 
a strongly interacting fermion superfluid (SF) in an
optical lattice.
Upon imposing the optical lattice potential on a SF of
$^6$Li atoms with near-unitary scattering,
at a density of two atoms per lattice well, it was found
\cite{ketterle_sflattice06} that
the SF retained its
coherence at weak to moderate lattice depths as deduced from the interference 
peaks in the momentum distribution of Feshbach molecules.
Beyond a certain critical lattice depth, however,
the interference peaks disappeared, signalling a possible SF
to band insulator (BI) quantum phase transition. 
Motivated by these experiments, theoretical studies \cite{zhai07,nikolic07} 
have focussed 
on the BI states of strongly interacting fermions in 
deep lattices, and shown that such BIs can be unstable
to Cooper pair formation below a critical lattice depth.
Interestingly, a toy model for such an interaction dependent
SF-BI transition was proposed and studied
some years ago \cite{nozieres99} in the context of the cuprate
superconductors.

In this paper, we focus on the problem of understanding
the multiband SF, and SF-BI transition, 
of fermions in a three dimensional (3D) optical lattice, at a 
density of two fermions per well, using a mean field
{\em intraband pairing approximation} which 
is equivalent to Anderson's idea of pairing exact time-reversed 
eigenstates \cite{anderson59}. This is known to be an excellent 
approximation in the context of 
disordered solid-state superconductors \cite{inhomscs} and in metallic 
nanoparticles \cite{vondelft}, where pairing between
single-particle states which are not related by time-reversal symmetry
gets suppressed by quenched randomness.
In the present context, we argue that the intraband pairing approximation
provides an internally self-consistent mean-field approach
and that interband pairing is expected to be dynamically suppressed.
Using this theory we obtain several interesting results, summarized below. 

On the theoretical front:
(i) We point out that
there exists a significant  discrepancy between the
earlier theoretical estimates \cite{zhai07,nikolic07} of the 
critical lattice depth for the SF-BI transition,
and we provide a physical resolution of this discrepancy. (ii)
Our mean field approach provides a numerically tractable
route to understanding properties of the SF phase which
was not considered in earlier theoretical work \cite{zhai07,nikolic07}.
(iii) Finally, we show that there are a
number of ways in which the cold atom multiband SF is distinct
from multiband solid state superconductors. These differences arise from the fact
that in solid state systems there is a cutoff in the pairing interaction, set by 
the Debye frequency in a conventional superconductor
or by the lattice cutoff in model tight-binding
Hamiltonians, which is absent in cold atom SFs. More generally, this
implies that results obtained on such effective lattice models in the
context of cold atom Fermi superfluids must be
interpreted with caution.

On the experimental front:
(i) As seen from Fig.~1, we find that the SF-BI transition, at 
unitarity,
occurs at a critical lattice depth $V^{(crit)}_L \approx 4 E_R$, where
$E_R = \pi^2 \hbar^2/(2 m_a b_L^2)$ is the atomic recoil energy, with
$b_L=\lambda/2$ being the lattice spacing ($\lambda$ being the 
laser wavelength) and $m_a$ being the atom mass.
Our result is somewhat larger than the experimental estimate
\cite{ketterle_sflattice06} 
of $V^{(crit)}_L \approx 3 E_R$, but is reasonably close, given that our 
approach 
is a mean-field theory, which neglects fluctuations (but see below).  
Our result for $V^{(crit)}_L$ coincides with the result of Ref.~\cite{zhai07}
but follows from a set of internally consistent assumptions.
We also show why the result of Ref.~\cite{nikolic07}, while obtained
within a valid formalism, leads to a significant overestimate of the critical 
lattice depth for the SF-BI transition at mean field level.
(ii) Fig.~1 shows the uniform and modulated
components of the condensate density. While the uniform
component decreases
monotonically with the lattice potential, the different modulated
components are nonmonotonic since they are
induced by the lattice potential. The maximum
value of the first order peak occurs at about half
the critical lattice depth and the peak value is about a factor of five
smaller than the uniform component, which is in a good agreement with 
the experimental data \cite{ketterle_sflattice06}.
(iii) As seen from Fig.~2, the 
excitation gap for fermionic quasiparticles is always nonzero at low
temperature, both in the SF and in the BI with the minimum
excitation gap occurring for lattice depths slightly smaller than the critical lattice depth.

The paper is organized as follows. Section \ref{sec:mft} contains a theoretical
discussion where we develop our mean-field 
theory and show that the intraband pairing ansatz gives an internally consistent 
description of the physics at density of two atoms per well. 
Section \ref{sec:results} explores the experimental consequences of
our mean-field theory. Specifically, we present results for the
critical lattice depth for the SF-BI transition, the dependence of both 
uniform and modulated components of the condensate density in the SF phase 
on the lattice depth and compare these results with the experiment. 
We also compute the quasiparticle excitation gap as a function of the lattice potential depth
and find a nonmonotonic dependence which can be tested in future experiments. 
We conclude in Section \ref{sec:conclusions} with a brief summary of the results.

\section{Model and Mean Field Theory}
\label{sec:mft}
Let us consider fermions moving in a 3D lattice potential
\be
\label{eq:1}
V(\bR)\! =\! - \frac{V_{L}}{2} \left[
\cos(\frac{2 \pi X}{b_L})\!+\!\cos(\frac{2 \pi Y}{b_L})
\!+\!\cos(\frac{2 \pi Z}{b_L}) \right],
\ee
with lattice constant $b_L=\lambda/2$, formed by three pairs
of counterpropagating laser beams with wavelength $\lambda$.
Let us use units where we measure momentum 
(distance) in units of $b^{-1}_L$ ($b_L$) and energies in 
units of the single atom recoil energy, $E_R = \pi^2 \hbar^2/2 
m_a b_L^2$.
Solving the (dimensionless) single-particle Schrodinger 
equation,
we find band energies $\varepsilon_{n \bk}$ and the Bloch
functions $u_{n \bk}(\br)$. Expanding the fermion field operator
in the Bloch basis as $\psi(\br) = \frac{1}{\sqrt{N}} \sum_{n \bk} 
\psi_{n \bk} u_{n \bk}(\br) {\rm e}^{i \bk\cdot\br}$, where $N$ is the total number of unit cells, 
and including the interactions, leads to the Hamiltonian
\be
\label{eq:2}
H = \sum_{n \bk \sigma} \xi_{n \bk}
\psi^\dg_{n \bk \sigma} \psi^\pdg_{n \bk \sigma} - w_S \int d^3\br 
\psi^\dg_{\upa}(\br) \psi^\dg_{\dna}(\br) \psi^\pdg_{\dna}(\br) 
\psi^\pdg_{\upa}(\br).
\ee
Here $\xi_{n \bk} = \varepsilon_{n \bk} - \mu$ 
($\mu$ is the chemical potential), $w_S = g_S/(b_L^3 E_R)$ is 
the (dimensionless)
contact interaction which can be tuned with a magnetic field,
and $g_S$ determines the s-wave scattering
length, $a_S$, via
\be
\label{eq:3}
\frac{m_a}{4\pi\hbar^2 a_S} 
= - \frac{1}{g_S(\Lambda)} + \int_0^\Lambda \frac{d^3\bQ}{(2 \pi)^3} \frac{m_a}{\hbar^2 \bQ^2}.
\ee
As indicated, $g_S$ must depend on the
ultraviolet cutoff $\Lambda$ to recover the measured (low 
energy) scattering length $a_S$. We ensure that $\Lambda$ is large
enough for physical properties to have converged to a cutoff-independent 
value in our numerical calculations. An important point to note is that
$g_S$ (and hence $w_S$) scales as $\sim 1/\Lambda$ when
$\Lambda \to \infty$.

To proceed with the many-body problem, we first follow the standard 
Bogoliubov-de Gennes method \cite{de Gennes} and
introduce mean fields
$\rho(\br) = 
\sum_\sigma \la \psi^\dg_{\sigma}(\br) \psi^\pdg_{\sigma}(\br) \ra$
and 
$\Delta(\br)= w_S \la \psi^\pdg_{\dna}(\br) \psi^\pdg_{\upa}(\br) \ra$,
where the expectation values are evaluated in the 
mean-field ground state which is to be determined self-consistently. 
Previous work \cite{zhai07,nikolic07} ignored the Hartree mean 
field $\rho({\bf r})$; we first show that this is justified in the cold
atom superfluid despite the
inhomogeneous density induced by the optical lattice. In order to
see this, we note that the inhomogeneous Hartree shift is simply
a renormalization of the optical lattice potential as $V_{eff}(\br) 
= V_L(\br) - w_S \rho(\br)/2$. In the limit that the cutoff $\Lambda \to \infty$,
$w_S \sim 1/\Lambda$, while the density $\rho(\br)$ is certainly
finite everywhere. This means that $V_{eff}(\br) = V_L(\br)$ and that
the Hartree terms play no role in the limit $\Lambda \to \infty$. We will
later see that this same argument fails for the pairing potential
$\Delta(\br)$ since $\la \psi^\pdg_{\dna}(\br) \psi^\pdg_{\upa}(\br) \ra$
is in fact ultraviolet divergent unlike $\rho(\br)$; this divergence
is precisely compensated by $w_S \sim 1/\Lambda$.

Let us proceed by introducing Fourier modes for the SF order parameter 
modulations as $\Delta(\br) = \sum_{\bg} \Delta(\bg) {\rm e}^{i\bg\cdot\br}$, as
well as for the Bloch functions,
$u_{n \bk }(\br) = \sum_{\bg} u_{n \bk}(\bg) {\rm e}^{i\bg\cdot\br}$, 
where $\bg$ are reciprocal lattice vectors.
$u_{n \bk}(\bg)$ can be chosen to be real due to time reversal symmetry.
We can then define matrix elements
\bea
\label{eq:4}
\Delta_{n n' \bk} &\equiv& \sum_{\bg} \Delta(\bg) F_{n n'  \bk}(\bg), 
\eea
where 
$F_{n n' \bk}(\bg) = \sum_{\bg'} u_{n \bk}(\bg + \bg') u_{n' \bk}(\bg')$.
The pairing matrix element $\Delta_{n n' \bk}$ corresponds to singlet pairing of
fermions at momenta $(\bk,-\bk)$. This leads to {\em intraband}
pairing for $n=n'$ and {\em interband} pairing for $n \neq n'$.
Both types of pairing matrix elements are generally nonzero. 
A mean field decoupling of the interaction term in $H$ leads to
\bea
\label{eq:6}
\!\!&\!\!H_{mf}&\!\!=\!\!\sum_{n \bk \sigma} \xi_{n \bk} \psi^\dg_{n \bk \sigma} 
\psi^\pdg_{n \bk \sigma} \nonumber \\
&-&\sum_{n n' \bk} \left[ \! \Delta_{n n' \bk} \psi^\dg_{n \bk \upa} \psi^\dg_{n' -\bk \dna} \!+\!
\Delta^*_{n n' \bk} \psi_{n' -\bk \dna} \psi_{n \bk \upa} \! \right]. 
\eea
The self-consistent mean field state can be obtained iteratively,
by diagonalizing $H_{mf}$
in Eq.~(\ref{eq:6}) and using those eigenstates to evaluate
the pairing field matrix 
elements in Eq.~(\ref{eq:4}) which yields the
new $H_{mf}$. This
is an extremely
computationally demanding task due to a large 
number of bands ($\gtrsim 100$) one needs to keep in order to converge to 
cutoff-independent 
results at unitarity. However, if one is interested in the critical 
lattice depth,
$V_L^{(crit)}$, at which the SF to BI transition occurs, the 
computations can be significantly simplified by studying the
linearized Bogoliubov-de Gennes equations,
\be
\label{eq:7}
\Delta(\bg) = \sum_{\bg'} {\cal K}(\bg, \bg') \Delta(\bg'), 
\ee
where the pairing kernel is given by
\bea
\label{eq:8}
{\cal K}(\bg, \bg')&=& \frac{w_S}{N} \sum_{n n' \bk}\frac{1 - n_F( \xi_{n \bk}) - 
n_F(\xi_{n' \bk})}{\xi_{n \bk} + \xi_{n' \bk}} \nonumber \\
&\times& F_{n n' \bk}(\bg) F_{n n' \bk}(\bg').
\eea
Here $n_F$ is the Fermi distribution function. (Ref.\cite{zhai07} incorrectly
assumed that this pairing kernel was diagonal in $\bg$.)
One can calculate $V^{(crit)}_L$ by diagonalizing the 
matrix $\delta_{\bg,\bg'} - {\cal K}(\bg, \bg')$, as was done in Ref.\cite{nikolic07},
and finding the value 
of $V_L$ at which its lowest eigenvalue becomes negative. This was
shown \cite{nikolic07} to lead to $V^{(crit)}_L \approx 45 E_R$ which
is more than an order of magnitude larger than the experimental
estimate $V^{(crit)}_L \approx 3 E_R$.

The reason for this enormous discrepancy can be 
understood if we look at the 
real-space structure of $\Delta(\br)$, corresponding to the lowest
eigenvalue of the pairing 
kernel ${\cal K}(\bg,\bg')$ near $V^{(crit)}_L$, by Fourier 
transforming the corresponding eigenvector.
We find that $\Delta(\br)$ near $V^{(crit)}_L$
is {\it very strongly modulated}; it is large near the center of each 
well of the optical lattice but is nearly zero in the region
between adjacent wells.
For a filling of two particles per well ($\bar{\rho}=2$), this
means that $\Delta(\br)$ varies rapidly on the scale of the
interparticle spacing $\sim 1/k_F \sim b_L$ which is comparable to 
the coherence length at unitarity. Amplitude fluctuations of the
order parameter will therefore be very important for such rapidly
varying components of $\Delta(\br)$ and will tend to suppress
such fast modulations. Since
interband pairing matrix elements only arise from $\Delta(\bg \neq 0)$,
this also means that the amplitude fluctuations will predominantly suppress 
interband pairing amplitudes. In short, a {\it mean-field} treatment of interband
pairing is internally inconsistent 
at the atom density of two atoms per well: it predicts a significant variation 
of the pairing mean field on short length scales, where amplitude fluctuations 
are expected to be significant. We do expect such a mean field theory to 
provide a better starting
point in the case where there is a large number of atoms in each
well so that amplitude fluctuations on the scale of $b_L$ can be expected 
to be small. In this case, the dominant fluctuations will be
quantum phase fluctuations which can drive a SF-BI transition
analogous to the
superconductor-insulator transition in Josephson junction arrays \cite{fazio}.
We will confine our attention here to a density 
of two atoms per unit cell. 

In contrast to interband pairing, the intraband pairing
is stronger when the mean-field potentials 
are more homogeneous, as seen from Eq.(\ref{eq:4}). 
Thus one can obtain a self-consistent mean-field description 
of the SF-BI transition in our system by just retaining intraband 
matrix elements in Eq.(\ref{eq:6}). This corresponds to Anderson's idea
of pairing time-reversed eigenstates \cite{anderson59}. 
This is known to be an excellent approximation in 
the context of the ordinary solid-state-based inhomogeneous, in particularly 
disordered, superconductors \cite{inhomscs} and in metallic nanoparticles \cite{vondelft}.
 In this case pairing between non-time-reversed 
eigenstates is suppressed by the randomness of the corresponding 
matrix elements. In our case the suppression is dynamical leading, 
however, to the 
same final result.  

Under this intraband approximation, the mean-field
Hamiltonian is obtained by setting 
$\Delta_{n n' \bk} = \Delta_{n \bk} \delta_{n,n'}$
in Eq.~(5).
Defining the Bogoliubov
quasiparticle dispersion as $\cE_{n \bk}
= \sqrt{\xi_{n \bk}^2 + \Delta^2_{n \bk}}$,
we are led to the mean-field equations
\bea
\label{eq:10}
\Delta(\bg)&\!\!=\!\!&
\frac{w_S}{N} \sum_{n \bk}
\frac{\Delta_{n \bk}}{2 {\cE}_{n \bk}}
\tanh\left(\frac{\cE_{n \bk}}{2 T}\right) F_{n n \bk}(\bg), \nonumber \\
\bar{\rho} &\!\!=\!\!& \frac{1}{N} \sum_{n \bk} \left[
1 - \frac{\xi_{n \bk}} {{\cE}_{n \bk}}
\tanh\left(\frac{\cE_{n \bk}}{2 T}\right) \right] ,
\eea
where $\bar \rho=2$ is the atom density per well. 
The pairing matrix elements are in turn determined via
$\Delta_{n \bk} = \sum_{\bg} \Delta(\bg) F_{n n \bk}(\bg)$.
In order to solve these equations we need to restrict
the number of bands. This is done by assuming $|\bg| < 2\pi\Lambda$, 
where $\Lambda$ is the cutoff in Eq.~(\ref{eq:3}). 
\section{Results}
\label{sec:results}
We now turn to some of the key results obtained from solving the intraband 
mean-field equations that we motivated 
in the previous section. We begin by pointing out some general qualitative
ways in which cold atom superfluids differ from ordinary solid state systems
although the mean field gap and number equations appear to be similar for
these two systems. We then turn to results obtained from numerically solving
the mean field equations for some experimentally relevant observables.

\subsection{What is special about atomic Fermi superfluids?}
We begin by 
pointing out an interesting and unexpected consequence of the intraband 
pairing ansatz --- namely, 
the pairing field in real space is completely uniform when we take
the limit $\Lambda \to \infty$ as we should. This can be seen by 
inspection of the expression for the Fourier component of the pairing 
field at wavevector $\bg$ in Eq.~(\ref{eq:10}). As seen from Eq.~(3),
the coupling constant $w_S$ vanishes as $\sim 1/\Lambda$ as the
cutoff is taken to infinity. This vanishing of the coupling constant
is cancelled by the expansion of phase space for pairing, i.e. the
number of bands at high energy which contribute to the sum in Eq.~(\ref{eq:10}), 
provided the matrix element $F_{nn\bk}(\bg)$ allows it. However,
high energy ($\varepsilon_{n \bk} \gg V_L$) states are simply free-particle 
states for which $F_{nn\bk}(\bg) \approx \delta_{\bg,0}$.
This means that only $\Delta(\bg=0)$ component of the pairing field 
survives in the limit $\Lambda \to \infty$. 
Note that this result is independent of the form of the lattice 
potential and can be expected to hold, for example, in the case of the random
optical lattice potential, which can be created by laser speckle 
\cite{Modugno}. This has potentially interesting implications
for the disorder-driven SF to insulator transition in cold atom SFs.    
Also note that these observations apply only to the cold atom SFs;
in conventional superconductors, the finite cutoff (at the Debye frequency in
conventional superconductors, or at the lattice cutoff in model tight binding
Hamiltonians)
does allow for spatial variations of $\Delta(\br)$
within the approximation of pairing time-reversed
eigenstates \cite{inhomscs}.
In terms of a Landau theory, if we have a uniform superfluid
order parameter and a periodic density modulation, then a symmetry
allowed term in the Landau functional such as $\int d^3\br
\rho(\br) |\Delta(\br)|^2$ will certainly lead to an order parameter modulation
with the same period as the density modulation. Our results indicate that
the coefficient of this term must vanish, in the intraband pairing approximation,
as the cutoff $\Lambda \to \infty$.
One final peculiarity of cold atom SFs is
that the pairing field (and the pairing gap) is completely {\em rigid}, i.e. 
it does not depend 
on the band index $n$ or the momentum $\bk$, unlike in multiband
solid state superconductors.
To the extent that interband pairing is suppressed, as we have argued
is the case for two atoms per well,
our above observations imply that any attempt
to model this multiband cold atom SF by imposing a finite band 
cutoff, as is customarily done in deriving
effective tight-binding lattice models, will generally lead to incorrect 
results.
We now turn to results obtained from a numerical solution of the intraband 
mean-field equations for observables of experimental interest.

\subsection{Critical lattice depth for the SF-BI transition}
We begin by investigating the critical lattice depth for the
SF-BI transition at atom density $\bar{\rho}=2$. 
For fermions with unitary scattering, 
the SF order parameter vanishes at $V^{(crit)}_L \approx 4 E_R$,
close to the value given in Ref.~\cite{zhai07} obtained by simply
assuming, in effect, that the order parameter is uniform. The results
and arguments presented in this paper justify such an assumption.
As we already mentioned, experimentally it appears that $V^{(crit)}_L 
\approx 3 E_R$ (note that
Chin {\it et al.} \cite{ketterle_sflattice06} quote their results in 
units of the molecular recoil $E_R/2$).
Part of the reason for the remaining relatively mild discrepancy with the 
experiment
might be quantum and thermal fluctuations beyond mean-field theory,
which are known to be somewhat important at unitarity even in the absence 
of the lattice
\cite{sensarma08}. One must also keep in mind, however, that
there is no good experimental estimate of the temperature after
the lattice is turned on \cite{ketterle_sflattice06}. 

\begin{figure}[t]
\includegraphics[width=8cm]{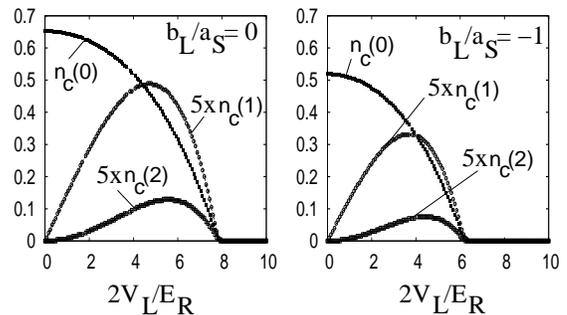}
\caption{Lattice depth dependence of the condensate density in the 
SF phase for different scattering lengths. 
$n_c(0)$ denotes the uniform component of the condensate density while
$n_c(1)$ and $n_c(2)$ correspond to modulated components 
at wavevectors $\bg=[2\pi,0,0]$ and $\bg=[2\pi,2\pi,0]$ respectively.}
\label{fig:deltaG}
\end{figure}    

\subsection{Condensate density}
It is important to realize that uniformity of the pairing field does not 
imply uniformity of the condensate density, which is what was measured in the 
experiment of Ref. \cite{ketterle_sflattice06}. The condensate density is
obtained from the two particle density matrix \cite{pairdensity}.
In a lattice environment, this condensate density has Fourier components
\be
\label{eq:11}
n_{c}(\bg) = \frac{1}{N} \sum_{n \bk} F_{n n \bk}(\bg) \left| \la \psi^\dg_{n \bk \upa} \psi^\dg_{n -\bk \dna}
\ra \right|^2. 
\ee
The sum over the band index $n$ in Eq.(\ref{eq:11}) is well defined in the limit $\Lambda 
\rightarrow \infty$ and all Fourier components of $n_{c}$ are in general 
nonzero. As shown in Fig.~1, the uniform component $n_{c}(\bg=0)$
decreases monotonically with increasing lattice depth, while the modulated
components have a nonmonotonic dependence on $V_L$ as observed in the
experiment. The ratio of the leading modulated component to the uniform
condensate density is in a good agreement with the data.

\subsection{Quasiparticle gap}
The SF state as well as the insulating state have a gap to
quasiparticle excitations. On the SF side, the quasiparticles
are Bogoliubov quasiparticles while the quasiparticles in the
insulating phase correspond to the original fermions.
As shown in Fig.~2, the minimum excitation
gap, $2 E_g$, in the SF phase decreases with $V_L$ (tracking
$2 \Delta(0)$ over a wide range of lattice depths) whereas it increases with
$V_L$ on the insulating side. This leads to a minimum in the
excitation gap for lattice depths close to but below $V^{(crit)}_L$. This
nonmonotonic variation of the gap, shown in Fig.~2, could
be tested in experiment. Note that there is a small window of
lattice depths, close to but below $V^{(crit)}_L$, where the quasiparticle
gap is slightly less than the gap of the underlying BI. This arises
because of particle-hole symmetry breaking and would not occur 
in particle-hole symmetric toy models \cite{nozieres99}.

\begin{figure}[t]
\includegraphics[width=7cm]{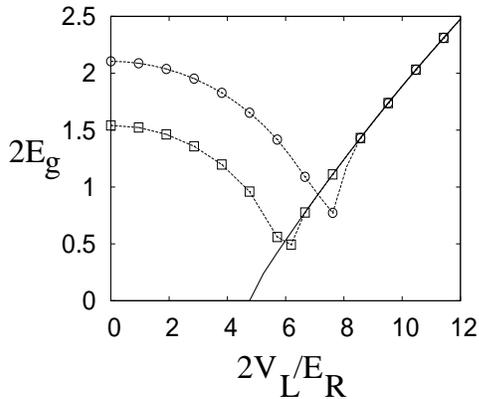}
\caption{Minimum excitation gap, $2 E_g$, as a function of lattice depth
in the SF and BI phases at low temperature for
scattering lengths corresponding to unitarity (open circles) and
for $a_S/b_L=-1$ (open squares). Solid line indicates the insulating
gap of the band problem. This gap corresponds to what would be measured in
typical experiments
probing the gap which will create two Bogoliubov quasiparticles in the SF 
or a particle-hole pair in the BI.}
\label{fig:gap}
\end{figure}

\section{Conclusions}
\label{sec:conclusions}
In this paper, we have discussed
the multiband SF phase of strongly interacting fermions
in an optical lattice. Our results for the critical lattice depth of the
SF-BI transition and the dependence of the Fourier
components of the condensate
density on the lattice depth are in a good agreement with
experimental data of Ref.~\cite{ketterle_sflattice06}. Our prediction for the quasiparticle gap in the
SF and insulating phases could be tested in future experiments.
We have also obtained new results of general relevance to cold atom SFs. 
In particular, the uniformity of the pairing amplitude under time-reversed-eigenstate 
pairing conditions is an interesting and surprising result which can be expected to have 
a strong effect, for example,  on the nature of the 
disorder-driven SF to insulator transition in cold atom SFs.  This will 
be explored in future work.  
Other directions for future research include a 
study of quantum fluctuations in the multiband SF, dynamical instabilities of 
such fermion SFs \cite{burkov08}, 
and the study of SF-insulator 
transitions in other lattice geometries \cite{zhao06}.

{\bf  Acknowledgments:}
We thank V. Galitski, H.R. Krishnamurthy, H. Moritz, P. Nikolic and
M. Randeria for 
useful comments and discussions.
We acknowledge support from the Sloan Foundation (A.P.), the Connaught
Fund (A.P.), and NSERC of Canada (A.A.B., A.P.).

\end{document}